\documentstyle[preprint,aps]{revtex}

\begin{document}
\draft
\preprint{IUCAA-51/2000}
\title{On a Peculiar Family of Static, Axisymmetric, Vacuum Solutions of
the Einstein Equations.}
\author{N. Dadhich \footnote{e-mail: nkd@iucaa.ernet.in}}
\address{The Inter-University Center for Astronomy and Astrophysics\\
PO Box 4, Ganesh Khind, Pune-411 007, INDIA.}
\author{G. Date \footnote{e-mail: shyam@imsc.ernet.in}}
\address{The Institute of Mathematical Sciences,
CIT Campus, Chennai-600 113, INDIA.}
\maketitle
\begin{abstract} 
The Zipoy-Voorhees family of static, axisymmetric vacuum solutions forms 
an interesting family in that it contains the Schwarzschild black hole
excepting which all other members have naked singularity.
We analyze some properties of the region near singularity by studying a 
natural family of 2-surfaces. We establish that these have the topology of 
the 2-sphere by an application of the Gauss-Bonnet theorem. By computing 
the area, we establish that the singular region is `point-like'. Isometric
embedding of these surfaces in the three dimensional Euclidean space is used
to distinguish the two types of deviations from spherical symmetry.
\end{abstract}

\vskip 0.50cm


\narrowtext

\section{Introduction}

The Weyl class of static, axisymmetric vacuum solutions of Einstein equation 
is very well studied and a systematic procedure for constructing
solutions is available \cite{stephani}. This class of solutions can be 
looked upon as describing finite size static bodies with axial symmetry.
These admit a regular horizon only when the symmetry turns spherical and
the solution is the Schwarzschild black hole. Among these is a class of so
called `prolate spheroidal' solutions found by Zipoy \cite{zipoy}. A
special family of this class is the subject of this note. \\

Solutions of this family are parameterized by two parameters, $m$ and
$A$. All are asymptotically flat and all except $A = 1$, which is the
Schwarzschild solution, have `naked singularity'. It is interesting to
note that arbitrarily small deviations from the Schwarzschild black hole
value of $A = 1$, which also
introduces deviations from sphericity, convert the event horizon
into an eternal naked (curvature) singularity. This makes the study of this
family interesting and this is undertaken here. \\

In section II, we sketch an alternative derivation of the solution.
Traditional approaches to construction of Weyl solutions have involved
use of `Newtonian potentials' corresponding to various source
configurations eg \cite{stephani,voorhees}. We discovered this solution
while studying the Kerr metric derivation given by Chandrasekhar
\cite{chandra} in some other context. The solution is arrived at essentially 
by  specializing Chandrasekhar's equations \cite{chandra} leading to the
Kerr solution, by switching off rotation ($\omega = 0$). This is the 
static limit of the Kerr geometry which would reduce to the
Schwarzschild black hole by the demand of the existence of a regular
horizon.\\

In section III, we analyze some properties of the solution. After
stating some elementary properties we focus on its $t = $constant, 
$r = $constant surfaces, $\Sigma_2$.  We analyze its topology and
intrinsic geometry and examine its `shape' as viewed in a three 
dimensional Euclidean space. It turns out that for the solutions for $A
> 1$ and $r$ very close to $2m$, these surfaces are {\em{not}}
completely embeddable in the Euclidean three dimensional space. This is
a rather unusual and interesting feature of this family of solutions.
What {\em{physical}} implication this feature has is at present an open 
question.\\

In the final section, we comment on possible physical implications of
the results and open issues. 
The notation and conventions used are those of Chandrasekhar
\cite{chandra} with the metric signature (+ - - -). 

\section{Derivation of The Solution}

The solution is fairly straightforward to derive. We begin by the
general equations given in \cite{chandra} for stationary, axisymmetric
space-times and specialize to the static form of the ansatz by setting
$\omega = 0$. The organization of equations and subsequent simplifications
is identical to that given in \cite{chandra}. Hence we give
only the key steps. \\

The metric is given by,

\begin{equation}
ds^2 = e^{2\nu} dt^2 - e^{2\psi} d\phi^2 - e^{2\mu_2} dr^2 - e^{2\mu_3}
d\theta^2.
\end{equation}

The metric coefficient functions depend only on $r, \theta$ and
$\Delta \equiv $exp$[2(\mu_3 - \mu_2)]$ is freely specifiable. Defining
$\beta \equiv \psi + \nu$, the vacuum equations become,

\begin{equation}
\partial_r( \sqrt{\Delta} \partial_r e^{\beta} ) + 
\partial_{\theta}( \frac{1}{\sqrt{\Delta}} \partial_{\theta} e^{\beta} ) 
\end{equation}
%
\begin{equation}
\partial_r \{ e^{\beta} \sqrt{\Delta} \partial_r (\psi - \nu) \} + 
\partial_{\theta} \{ \frac{e^{\beta}}{\sqrt{\Delta}} \partial_{\theta} (\psi
- \nu) \} 
\end{equation}
%
\begin{equation}
\partial_r\partial_{\theta} \beta - 
\partial_r \beta \partial_{\theta} \mu_2 - 
\partial_{\theta} \beta \partial_r \mu_3 +
\partial_r \nu \partial_{\theta} \nu +
\partial_r \psi \partial_{\theta} \psi 
\end{equation}
%
\begin{eqnarray}
4 \sqrt{\Delta} 
\{ \partial_r \beta \partial_r \mu_3 + 
\partial_r \psi \partial_r \nu \}  - \frac{4}{\sqrt{\Delta}}
\{ \partial_{\theta} \beta \partial_{\theta} \mu_2 + 
\partial_{\theta} \psi \partial_{\theta} \nu \}  \nonumber \\
= ~ 2 e^{-\beta} 
\{ \partial_r( \sqrt{\Delta} \partial_r e^{\beta} ) - 
\partial_{\theta}( \frac{1}{\sqrt{\Delta}} \partial_{\theta} e^{\beta} ) \}
\end{eqnarray}

In order to admit possibility of a horizon, following Chandrasekhar, we
set exp$\{ \beta(r, \theta) \} \equiv \sqrt{\Delta(r)} f( \theta )$.
Equation (2) then leads to $\Delta = r^2 + d_1 r + d_2 , f(\theta) =
sin(\theta)$. In anticipation of spherical topology we have required
$f(\theta)$ to vanish at the `poles' at $\theta = 0, \pi$. We will assume
that $d_i$ constants are such that $\Delta = 0$ could have two real
roots and we may write $\Delta(r) = (r - r_+)(r - r_-)$. The remaining
equations now become equations for $\nu$ and $\mu_2$ (say) by
eliminating $\psi$ in favour of $\nu, \beta$ and by eliminating $\mu_3$
in favour of $\mu_2, \Delta$. The class of solutions we are interested in
is obtained by assuming that $\nu(r, \theta)$ {\it{ is independent of 
$\theta$}}. The equations simplify further and lead to the following 
solution: 

\begin{eqnarray}  
\nu & = & \frac{C}{r_+ - r_-} \ell n\{ \frac{r - r_+}{r - r_-} \} + D
~~~~~~~~~~,~~~~~~~~~~ \bar{C} \equiv C^2 - \frac{(r_+ - r_-)^2}{4} ;
\nonumber \\ 
\mu_2 & = & -\frac{C}{r_+ - r_-} \ell n\{ \frac{r - r_+}{r - r_-} \} - 
\frac{2\bar{C}}{(r_+ - r_-)^2} \ell n\{ \frac{sin^2\theta}{\Delta} +
\frac{4}{(r_+ - r_-)^2} \} + \Phi_0 \nonumber \\
\mu_3 & = & \mu_2 + \frac{1}{2} \ell n \Delta 
~~~~~~~~~~~~~~~~~~~~~~~~~~~~,~~~~~~~~~~ \Delta = (r - r_+)(r - r_-) \nonumber \\
\beta & = & \ell n ( sin(\theta) ) + \frac{1}{2} \ell n ( \Delta )
~~~~~~~~~~~~~~~,~~~~~~~~~~ \psi ~~ = ~~ \beta - \nu 
\end{eqnarray}  

Here $C, D$ and $\Phi_0$ (and of course $r_\pm$) are constants of integration. 
By redefining $r \rightarrow r + r_- $ one sees that $r_- $ now appears
only in the combination $r_+ - r_- \equiv 2m \ge 0$. For vacuum
solution, the metric is always defined up to an over all multiplicative
constant. We also have the freedom to redefine $t$ and $\phi$. Using
these one can effectively absorb the constants $D, \Phi_0$. Trading $C$
for $mA$, the final form of the metric can be written as:

\begin{eqnarray}
ds^2 & = & F^A dt^2 - F^{-A} G^{-B} dr^2 - \Delta F^{-A} G^{-B} d\theta^2 -
\Delta F^{-A} sin^2(\theta) d\phi^2  \\
\mbox{where,} ~~
\Delta & = & r^2 - 2mr ~~,~~ F = \frac{\Delta}{r^2} 
~~,~~ G = 1 + \frac{m^2}{\Delta} sin^2(\theta) ~~\mbox{and}~~ 
B \equiv A^2 - 1  \nonumber
\end{eqnarray}

Let us quickly remark that the solution can be expressed in a standard
form \cite{stephani} by the following identifications:

\begin{equation}
\rho ~ = ~ \sqrt{\Delta} sin \theta ~~,~~ z ~ = ~ 
\frac{1}{2} \frac{d\Delta}{dr} cos \theta ~~,~~ e^{2U} ~ = ~ 
F^A ~~\mbox{and}~~ e^{2k} ~ = ~ G^{-A^2} .
\end{equation}

So that the metric is given by,

\begin{equation}
ds^2 ~ = ~ e^{2U} dt^2 - \rho^2 e^{-2U} d\phi^2 - e^{-2U + 2k} \left(
d\rho^2 + dz^2 \right)
\end{equation}

We also note that this solution corresponds to the `prolate' family of
solutions found by Zipoy \cite{zipoy}. In equations (43-45) of \cite{zipoy},
put $\gamma = 0$. A trivial shift of his $r$ gives the above solution
with his $\beta$ same as our $A$. Voorhees \cite{voorhees} also obtained
the solution as an illustration of his method of constructing Weyl
solutions. He also elaborated some of its properties. \\

In the next section we recall some of the basic properties and then focus 
on the $t =$ constant and $r = $ constant surfaces $\Sigma_2$. 

\section{Properties of The Solution}

We have a two parameter family of static, axisymmetric solutions
potentially containing a horizon. For $A = 1$ we obtain the Schwarzschild
solution with mass $m$, $r$ being the usual areal radial coordinate and
$r = 2m$ being the horizon. The spherical symmetry is restored for this
value. For large $r$, the behaviour of the solution shows that it is
asymptotically flat with $mA$ being the ADM mass. We are primarily
interested in parameter values closer to the Schwarzschild case. Thus we
take $m > 0$ and $A > 0$ so that the mass is also positive ($m = 0$
gives Minkowski space-time). We now concentrate on $A \ne 1$. \\

To locate possible singular regions, we compute curvature components. 
For example (in Chandrasekhar's notation), 
\begin{equation}
R_{0101} = -A G^B \Delta^{A -2} r^{-2A} mr (1 - \frac{(A + 1) m}{r})
\end{equation}

As $r \rightarrow 2m$, $\Delta \rightarrow 0$. For $A \ne 1 ~ (B \ne 0) $, 
the net power of $\Delta$ becomes $A - A^2 -1$ which is 
always negative for all $real ~ A$. Thus we get a curvature singularity as
$\Delta \rightarrow 0$.
Other curvature components and Kretschmann scalar show similar behaviour.
For $A = 1$, there is no contribution from $G$ and the curvature
component reduces to $-\frac{m}{r^3}$. Since for $r \rightarrow 2m$
itself we encounter curvature singularity, there is no point in
exploring $r < 2m$ region. Since there is no other $r > 2m$ where the norm
of the stationary Killing vector ($g_{tt}$) can vanish, the singularity
must be naked. Thus except for $A = 1$ where one has a non singular
Killing horizon, for all other values of $A$ the space-time has naked
singularity (diverging curvatures). This is an eternal
naked singularity in an asymptotically flat space-time with {\em{positive}}
ADM mass. From now on we limit ourselves to the
non-singular coordinate range $r > 2m$. In order to gain some intuitive
understanding of the region near the singularity as well as to see how
the $A \ne 0$ `effects' are seen in large $r$ regions, we focus now on the
$t = $ constant and $r = $ constant surfaces, generically denoted by 
$\Sigma_2$. This naturally leads us to study the intrinsic geometry of 
$\Sigma_2$. \\

The metric as given has all the coordinates as local coordinates. We
have implicitly assumed the $\theta, \phi$ to be the standard spherical
polar angles on $S^2$. As such the non-singular form of the metric is
not directly in conflict with such an interpretation of the coordinates.
(While $\phi$ can be taken as azimuthal angle referring to the
axisymmetry, interpretation of $\theta$ as the polar angle is not
automatic.) One way to check that $\Sigma_2$ indeed can have spherical
topology is to verify the Gauss-Bonnet theorem. This is precisely what
is done below. \\

It is straight forward to compute the Ricci scalar for the intrinsic metric on
a $\Sigma_2$. Since $r$ is a constant, defining $\alpha \equiv \Delta
F^{-A}$ and $\beta \equiv \frac{m^2}{\Delta}$ one obtains the Ricci
scalar as,

\begin{eqnarray}
R(g) & = & - \frac{1}{2} \left[ \partial_{\theta} \left(
\frac{\partial_{\theta} g_{\phi\phi}}{det(g)} \right) +
\frac{1}{det(g)}\partial_{\theta}^2 g_{\phi\phi} \right] \\
& = & \frac{2}{\alpha} \{ 1 + \beta sin^2\theta\}^{(B -1)} \{
(\beta B - 1) - \beta (1 + B) sin^2\theta \}
\end{eqnarray}

If topology of $\Sigma_2$ is indeed $S^2$, then integral of
$\sqrt{det(g)} R(g)$ over the full range of the coordinates should equal
$8\pi$. In terms of $x = cos\theta$, the integral can be expressed as,

\begin{equation}
\frac{1}{8\pi} \int_{\Sigma_2} \sqrt{det(g)} R(g)  =  
\beta^{B/2} \int_0^1 dx ~~ \frac{\frac{1 + \beta}{\beta} - (B + 1) x^2}
{\left(\frac{1 + \beta}{\beta} - x^2 \right)^{1 - B/2}}
\end{equation}

Clearly for $B = 0$ (Schwarzschild), the right hand side is 1 as
expected. Defining $\xi \equiv (1 + \beta)/\beta $, we get,

\begin{eqnarray}
\frac{1}{8\pi} \int_{\Sigma_2} \sqrt{det(g)} R(g) 
& = & (\xi - 1)^{-B/2} \left[ (B + 1) - 2 \xi \partial_{\xi} \right] g(
\xi, B) \\
\mbox{where,} ~~~~~~~~
g(\xi, B) & \equiv & \int_0^1 dx ~ (\xi - x^2)^{B/2} \nonumber
\end{eqnarray}

Note that $\xi > 1$ and hence the function $g(\xi, B)$ is well defined
and is in fact differentiable. One can simplify the $\xi$ dependence by
defining $x = \sqrt{\xi} y$ in the integral. It is now easy to evaluate
the $\xi$-derivative and check explicitly that the right hand side
is indeed 1 for all values of $B$ and $\beta$. {\it{Thus topology of 
$\Sigma_2$ surfaces is indeed that of $S^2$.}} \\

Next we compute the areas of these spheres. This will allow us to
estimate the `size' of the region of high curvature. The area is given by,

\begin{equation}
\mbox{Area(r)} = 4\pi ~ r^{2A} ~ \Delta^{(1 - A)^2/2} ~ \left[ \int_0^1 dx ~
\{ \Delta + m^2 (1 - x^2) \}^{(1 - A^2)/2} \right]
\end{equation}

Putting $r = m(\mu + 1), x = \mu y$ and $\gamma = (1 - A^2)/2$, the
area can be expressed as,

\begin{equation}
\mbox{Area($\mu,A$)} = 4\pi ~ m^2 ~ \mu^{2 - A^2} ~ (\mu + 1)^{(1 + A)^2/2} ~ 
(\mu - 1)^{(1 - A)^2/2} ~~ \int_0^{1/\mu}~ dy ~ (1 - y^2)^{\gamma}
\end{equation}

Since $r > 2m$ we have $\mu > 1$ and manifestly $\gamma < 1/2$. The area is 
clearly well defined and is a positive, finite
quantity for all finite $\mu > 1$. Our interest is to estimate the
behaviour of the area as $\mu \rightarrow 1$. Once again for $A = 1$, we
recover the expected answer. \\

It is easy to see that for $0 < A < 2, ~ A \ne 1$, the area vanishes as $r
\rightarrow 2m$ while for larger values of $A$ the area blows up. The 
vanishing area indicates that the singular region is `point-like'. For
diverging area we can not say so. Since
we are mainly interested in values of $A$ near the Schwarzschild value,
we see that for these values the area vanishes and hence the curvature
singularity is `point-like'. The coordinate value $r = 2m$ really
corresponds to areal radial coordinate vanishing. For the subsequent
analysis we will limit ourselves to solutions with `point-like' singular
region, i.e. to $0 < A < 2$.\\

\underline{Note:} If we assume that the Ricci scalar satisfies $a \le R
\le b$ everywhere on a spherical surface for some constants $a, b$ 
depending on the surface, then the Gauss-Bonnet integral for spherical 
topology gives,

\begin{equation}
a ~ \mbox{(Area)} ~ \le 8 \pi ~ \le ~ b ~ \mbox{(Area)} 
\end{equation}

Thus if the area vanishes then the Ricci scalar must be unbounded
above while if the area diverges then either the Ricci scalar must vanish
in a precise manner (eg usual $S^2$ metric for large radius) or $a$ 
must be negative or $a$ can go to $-\infty$. One can check from our 
expressions that this is indeed the case. Observe that, the $\Delta
\rightarrow 0$ behaviour of the Ricci scalar is different along the
poles and along generic $\theta$ directions.\\

Having estimated the `size' of the singular region, we now look for its
`shape'. All our intuitive understanding of `shape' of an object is
based on its embedding in three dimensional {\it{Euclidean}} space. A
natural formulation of `shape' of $\Sigma_2$ is to look for its
embedding in the Euclidean space of dimension 3. The embedding is to be
such that the induced metric on the image is the same as the intrinsic
metric on it, in short, an isometric embedding. \\

Using the symmetries of the Euclidean space and choosing the axis of
symmetry to be the Z-axis, a natural ansatz for the embedding is:

\begin{equation}
X(\theta, \phi) ~ = ~ x(\theta)~cos\phi ~~~,~~~
Y(\theta, \phi) ~ = ~ y(\theta)~sin\phi ~~~,~~~
Z(\theta, \phi) ~ = ~ z(\theta) ~.
\end{equation}

The demand that the induced metric be axisymmetric ($\phi$-independent)
and diagonal, leads to $y(\theta) = \pm x(\theta)$ and without loss of
generality we choose the + sign. Equating with the intrinsic metric then
gives (prime denotes derivative with respect $\theta$),

\begin{equation}
x(\theta) = \pm \sqrt{\alpha} ~ sin\theta ~~~,~~~
(z^{\prime}) ~ = ~ \pm \sqrt{\alpha} ~ \left[ (1 + \beta ~ sin^2\theta)^{-B} -
cos^2\theta \right]^{1/2}
\end{equation}

Without loss of generality we can choose positive sign for $x(\theta)$ and
negative sign for $z(\theta)$ equations above.
The equation for $z(\theta)$ is an ordinary differential equation with
one constant of integration which corresponds to the choice of origin.
It is invariant under reflection about the equator.  We can thus solve 
the equation for $0 \le \theta \le \pi/2$. For convenience, let us denote
the expression in the square brackets in the above equation by 
$f(\theta, \beta, B)$. We are interested in the behaviour of this function
for $0 \le \theta \le \pi/2$, for $0 < \beta $ and for $-1 < B < 3$ 
($0 < A < 2$). \\

Clearly $f$ is non-negative for all $\beta, \theta$ if $B \le 0$. 
{\it{however for $B > 0$ and for some $\beta$, it can be
negative for $\theta$ near the poles, indicating failure of complete
embedding.}}\\

Computing the first derivative of $f$ with respect to $\theta$ one sees that
$\theta = 0, \pi/2$ are two extrema with a possibility of a third extremum
at some $\hat{\theta}$. This third extremum, if exists, is always a 
minimum. For $\beta B < 1$, the third extremum does {\em{not}} exist while for 
$\beta B = 1$, it coincides with $\theta = 0$. Then $f$ is non-negative 
for all $\theta$. Thus for $\beta B \le 1$,
complete embedding is possible. For $\beta B > 1$, the minimum at 
$\hat{\theta} > 0 $ exists and then $f$ is negative for $0 \le \theta \le 
\theta_{max}$. The $\theta_{max}$ is to be obtained by solving the 
transcendental equation $f = 0$. Thus for $\beta B > 1$, caps nears the 
poles are not embeddable. \\

The condition $\beta B > 1$ corresponds to $r$ near $2m$. Since for $B$ close 
to zero, $\beta$ is very large and thus corresponds to $r$ very close to
$2m$. The farthest $r$ would be for the largest $B ~(= A^2 -1) $ which 
for us is 3 ($A < 2$). This translates to the farthest $r$ being $3m$. 
Thus for $r > 3m$, we will always have complete embedding. Curiously,
$r = 3m$ happens to be the radius of the photon circular orbit for the
Schwarzschild black hole.\\ 

While precise shape for complete embedding can be plotted by numerically 
integrating the equation, that is however not our main concern. In general the
`shape' will be an ellipsoid. For $B$ close to zero, the ellipsoid will 
shrink as $\Delta \rightarrow 0$. \\


\section{Discussion}

One of the striking features of this family of solutions is that for a
slightest deviation from spherical symmetry, the horizon disappears and
instead a naked singularity appears which is eternal by virtue of the 
staticity of the space-time. The departure from sphericity, equivalently
from the Schwarzschild geometry, is indicated by the parameter $A \ne 1$.
Note that $A = +1$ corresponds to the Schwarzschild black hole 
while $A = -1$ corresponds to the negative mass Schwarzschild
solution. We are assuming of course that the parameter $m$ is positive.
In order to include the Schwarzschild black hole, we have chosen $A >
0$. 
The parameter value $A = 2$, demarcates whether the area of $\Sigma_2$ 
for $r$ near $2m$ vanishes ($A \le 2$) or diverges ($A > 2)$. 
Likewise, the Schwarzschild value $A = 1$ demarcates
whether the surface $\Sigma_2$ for $r$ near $2m$ is completely embeddable
($A \le 1$) or is {\em{partially}} embeddable ($A > 1$). \\

Of course it is more likely that the solutions are valid for $r$
sufficiently large and that a physical body will fill in the interior
region. This leads one to the question of a source configuration to
which the above family is an exterior space-time \cite{voorhees}. While 
it would be natural to expect that a two surface across which an interior
solution is matched to the above solution is indeed one of the $\Sigma_2$
surfaces, it is by no means obvious that it must be so. For instance one
could have a 2-surfaces defined by $r = r(\theta)$. The Killing norm
however will not be constant over such surfaces. For large $r$ values
this will correspond to the matching surface not having constant
Newtonian potential. One could of 
course analyze the shape of any such matching surfaces by similar 
techniques. One must emphasize that the Euclidean space used for the 
isometric embedding is {\em{not}} the physical space whose metric near 
the vicinity of the surface is not Euclidean. It has been used as a way to 
discriminate among different members of the family \footnote{While the
paper was being written, a preprint by N. Pelavas, N. Neary and K. Lake
appeared on the e-print archive which applies similar methods to study
properties of the instantaneous ergo surface of the Kerr metric, 
gr-qc/0012052.}. \\

These solutions could be viewed as a subset (static) of possible end 
states of a non-spherical collapse. Black hole uniqueness results would
indicate that if censorship holds then either spherical symmetry must 
be restored (staticity results) or rotation must set in (only 
stationarity results) during the collapse. For the possibility of static 
end state, it is only a single special value for which a naked singularity
is avoided. In the light of these observations, it would be exceedingly 
interesting to see what features emerge in an `axisymmetric' collapse 
with the above family being used for exterior matching just as the 
Schwarzschild is used in the study of spherical collapse. \\

{\underline{Acknowledgments:}} G. D. would like to acknowledge a
helpful e-mail correspondence with Kapil Paranjape of the Institute of 
Mathematical Sciences regarding the formulation of the notion of
`shape'. He would also like to thank IUCAA for the warm hospitality and
conducive atmosphere provided during his visit during which this work
was completed.

\end{document}